b%
\documentclass[aps,preprint,showpacs,showkeys,nofootinbib,floatfix,superscriptaddress]{revtex4}

\usepackage{graphicx}
\usepackage{amsmath}
\usepackage{amsfonts}
\usepackage{bm}

\newcommand{\bea}{\begin{eqnarray}}
\newcommand{\eea}{\end{eqnarray}}
\newcommand{\beq}{\begin{equation}}
\newcommand{\eeq}{\end{equation}}
\newcommand{\bqa}{\begin{eqnarray}}
\newcommand{\eqa}{\end{eqnarray}}

\newcommand{\qvec}{{\bf q}}

\def\mqo2{{\!\!\!}}

\begin{document}

\preprint{INT-PUB-10-020}
\preprint{HISKP-TH-10-11}
\title{Beyond Universality in Three-Body Recombination:\\
An Effective Field Theory treatment}
\author{Chen Ji}\email{jichen@phy.ohiou.edu}
\affiliation{Department of Physics and Astronomy, Ohio University, Athens, OH\ 45701, USA\\}
\author{Daniel Phillips}\email{phillips@phy.ohiou.edu}
\affiliation{Department of Physics and Astronomy, Ohio University, Athens, OH\ 45701, USA\\}
\affiliation{Helmholtz-Institut f\"ur Strahlen- und Kernphysik, Universit\"at Bonn, D-53115, Bonn, Germany}
\author{Lucas Platter}\email{lplatter@phys.washington.edu}
\affiliation{Department of Physics, Ohio State University, Columbus OH\ 43120, USA\\}
\affiliation{Institute for Nuclear Theory,
        University of Washington, Seattle, WA\ 98102, USA\\}

\date{\today}

\begin{abstract}
  We discuss the impact of a finite effective range on three-body
  systems interacting through a large two-body scattering length. By employing
  a perturbative analysis in an effective field theory well suited to this
  scale hierarchy we find that an additional three-body parameter is required
  for consistent renormalization once range corrections are considered. This
  allows us to extend previously discussed universal relations between
  different observables in the recombination of cold atoms to account for the
  presence of a finite effective range.  We show that such range corrections allow us to simultaneously describe the positive and
    negative scattering-length loss features observed in recombination with
    ${}^7$Li atoms by the Bar-Ilan group. They do not, however, significantly
    reduce the disagreement between the universal relations and the data of the 
Rice group on ${}^7$Li recombination at positive and negative scattering lengths.
\end{abstract}



\smallskip
\maketitle
\section{Introduction}
In the 1970s Vitaly Efimov showed that the non-relativistic three-body
system with two-body scattering length, $a$, much larger than the
range of the underlying interaction displays universal properties
that are independent of the details of the interaction \cite{Efimov70,
 Efimov73}.  In particular, in the unitary limit $|a| \rightarrow
\infty$, there exists a tower of three-body states ({\it trimers})
with a geometric spectrum,
\begin{eqnarray}
E^{(n)}_T  = (e^{-2\pi/s_0})^{n-n_*} \hbar^2 \kappa^2_* /m,
\label{kappa-star}
\end{eqnarray}
where $\kappa_*$ is the binding wavenumber of the trimer
labeled by $n_*$, $m$ is the mass of the particles, and $s_0=1.00624$.
Eq.~\eqref{kappa-star} and other universal predictions have received
significant attention recently in both atomic physics and nuclear
physics. In the latter case, the three-nucleon system and
certain halo nuclei display features reflecting their status as
(effective) three-body systems which are near the unitary limit (for
reviews see \cite{Braaten:2004rn,Platter:2009gz}).

In atomic physics, the three-body recombination rate of ultracold
atoms is an observable that displays the discrete-scale invariance of
Efimov systems: it is proportional to $a^4$ times a function that is
log-periodic in $a$ \cite{Braaten:2004rn}. Several recent experiments
have demonstrated the existence of such {\it Efimov physics} by
measuring the recombination rate as a function of the scattering
length. For example, Ref.~\cite{Gross:2009} measured the three-body
recombination rate for $^{7}$Li atoms. Gross {\it et al.} found a
recombination minimum at a scattering length $a \equiv a_{*
  0}\approx 1160 a_B$, as well as a rate enhancement when $a =
a'_* \approx -264(11) a_B$.  (Here and below $a_B$ denotes the
Bohr radius.)

The ratio $a'_*/a_{* 0}$ is very close to the ``universal"
result~\cite{Braaten:2004rn,Helfrich:2010yr}:
\begin{equation}
a'_*=-\exp\left[\frac{(2n+1) \pi}{2 s_0}\right] a_{*0} \stackrel{n=-1}{\longrightarrow} -0.21 a_{*0}.
\label{eq:ratio}
\end{equation}
However, the interactions between ${}^{7}$Li atoms have a natural, van der
Waals, length scale $\sim 100 a_B$, so significant corrections to this
universal prediction for $a_*'$ are expected. In particular, the
  impact of a finite effective range on the scattering-length dependence of
  Efimov physics has been discussed in two recent
  papers~\cite{Platter:2008cx,Thogerson:2008}.
  In contrast to those works,
here we include the effect of a finite effective range perturbatively and
analyze the consequences for universal predictions in the
three-body sector. We do this by setting up a perturbation theory around the
unitary limit, and thus organize the corrections that manifest the differences
between different large--scattering-length systems. These are effects beyond
universality, and their inclusion allows us to extend the reach of Efimov's
ideas. The small parameters in this effective field theory (EFT) expansion are
$\ell/|a|$ and $\ell k$, where $k$ denotes the momentum scale of the problem
under consideration, and is $\sim 1/|a|$ for the recombination processes that
are our concern here. This ``short-range EFT" applies to all non-relativistic
systems in which $|a|$ is much larger than the range of the underlying
interaction $\ell$. Part of the dynamics at scale $\ell$ enters
  observables via the two-body effective range, $r_s$, and can be
  straightforwardly accounted for. But, other effects due to the finite range
  of the inter-atomic force get encoded in new {\it three-body} EFT
  parameters. These must be included in the calculation if it is to contain
  all effects up to a given order in the EFT. In this way we systematically
  approximate the dynamics of any finite-range interaction that gives a large
  scattering length $a$.  The main result of our EFT analysis 
  at
next-to-leading order (NLO) 
is that, if (and only if) scattering-length-dependent observables are
considered, one such additional three-body parameter must be included in the
calculation in order to guarantee the accuracy of the EFT's
predictions. To demonstrate the implications and limitations of this
  result, we use this approach to 
  calculate the three-body recombination rate
of $^7$Li in the hyperfine states relevant to the experiments
performed by Pollack {\it et al.}  \cite{pollack:2009} and Gross {\it et al.}
\cite{Gross:2009}.

\section{Effective Field Theory}
EFTs are a standard tool for calculating low-energy observables in
systems with a separation of scales. Here we will consider EFT for
particles interacting solely via short-range interactions. Another
successful example of an EFT is chiral perturbation theory ($\chi$PT)
that identifies pions as the Goldstone bosons of low-energy QCD (see
Ref.~\cite{Bijnens:2006zp} for a recent review). At the heart of every
EFT is a Lagrangian, which contains all possible operators allowed by
the underlying symmetries. The short-range EFT applies to
non-relativistic particles with a large scattering length and contains
only contact interactions. In this EFT we have, at leading order in
an $r_s/|a|$ (or, equivalently for our purposes, $\ell/|a|$) expansion
\begin{equation}
 \mathcal{L}={\psi}^{\dagger}\left(i
 \partial_t - \frac{\overrightarrow{\nabla}^2}{2m}\right)\psi
-\frac{C_0}{2}(\psi^\dagger\psi)^2 - \frac{D_0}{6}(\psi^\dagger\psi)^3+\ldots,
\label{eq:5}
\end{equation}
with $\psi$ our matter fields, and $C_0$ a two-body parameter fixed by
the physical scale $a$.  At leading order the two-body amplitude is
obtained by summing all two-body diagrams that contain only the
four-boson operator proportional to $C_0$. Requiring that the
two-body amplitude has the physical scattering length
leads to a renormalization condition for 
$C_0$:
\begin{equation}
\label{eq:7}
\frac{1}{C_0}=\frac{m}{4 \pi a} - \frac{m \Lambda}{2 \pi^2},
\end{equation}
if cutoff regularization with a cutoff $\Lambda$ is employed. 

To go beyond leading order an ordering scheme derived from the
requirement that each new order of the calculation gives contributions
to observables 
that scale
with powers of $r_s/|a|$ or $k r_s$ is
required~\cite{Beane:2000fx,Bedaque:2002mn}. This extends
Eq.~(\ref{eq:5}) to a predictive framework in which a well-defined
class of terms has to be evaluated at every order in this
small-parameter expansion.


\section{The Three-Body System}
\label{sec:three-body-system}
In Refs.~\cite{Bedaque:1998kg,Bedaque:1998km} Bedaque {\it et al.}
showed that this EFT is well-suited to describing the three-body
problem near the unitary limit, and facilitates derivation of Efimov's
results. In particular, they showed that a three-body parameter $D_0$ must be present
at LO
in order to obtain renormalized quantities. 

This result was obtained by using the Lagrangian above to derive an
integral equation for atom-dimer scattering. The $S$-wave projected
amplitude, ${\mathcal A}_0$, for scattering from relative-momentum
state $k$ into relative-momentum state $p$, at energy $E$, is given by
\begin{eqnarray}
{\mathcal A}_0 (p, k; E)  &=&
\frac{8 \pi}{a p k} 
\left[
\ln \left(\frac{p^2 + pk + k^2 -mE}
      {p^2 - pk + k^2 - mE}\right)+\frac{2H_0(\Lambda) p k}{\Lambda^2}
\right]
\nonumber
\\
&&\hspace{-2.5cm}+
\frac{2}{\pi} \int_0^\Lambda dq \, \frac{q}{p} 
\left[\ln \left(\frac{p^2 +pq + q^2 - mE}
            {p^2 - pq + q^2 -mE}\right)+\frac{2H_0(\Lambda) p q}{\Lambda^2}\right] \,
\frac{ {\mathcal A}_0 (q, k; E)}
{- 1/a + \sqrt{3q^2/4 -mE -i \epsilon} }
\label{BHvK}
\end{eqnarray}
with $\epsilon$ a positive infinitesimal. 
This equation (with $H_0=0$) is known as the
Skorniakov--Ter-Martirosian (STM) equation~\cite{STM57}.  The
additional term $\sim H_0(\Lambda)$ can be thought of as a
three-body force. It is proportional to $D_0/C_0^2$, and in general is
a function of the integral-equation cutoff.  The value of $H_0$ is
fixed by reproducing the value of one three-body observable, e.g. the
binding energy of a particular three-body bound state at some fixed
scattering length. This relates $H_0$ to a  physical scale in the three-body system.
A common choice for this scale is $\kappa_*$, which can be thought of
as the binding momentum of one three-body bound state in the
limit $|a| \rightarrow \infty$.  Once $\kappa_*$---or
equivalently $H_0$---is fixed, LO predictions follow for all other
three-body observables, including observables at other scattering
lengths which satisfy $|a| \gg \ell$.

For $p |a|, p/k \gg 1$, the asymptotic form of the amplitude
$\mathcal{A}_0(p,k)$ is $\sim p^{\pm i s_0-1}$.  Corrections are
suppressed by powers of $1/(|a| p)$ or $\ell p$~\cite{Bedaque:2002yg}:
\begin{equation}
 \label{eq:1}
 \mathcal{A}_0(p,k) 
 =\mathcal{N}(k/\Lambda)\biggl\{\frac{1}{p}\sin\left[s_0\ln\left(\frac{p}{\kappa_*}\right)\right]
+\frac{8
 |\mathcal{C}_{-1}|}{a p^2}\sin\left[s_0\ln\left(\frac{p}{\kappa_*}\right)+\arg(\mathcal{C}_{-1})\right]
+...\biggr\}~.
\end{equation}
Here $\mathcal{N}(k/\Lambda)$ denotes a normalization factor and
$\mathcal{C}_{-1}$ is a complex number that can
be obtained from the Mellin transform of the asymptotic form of the
kernel of the integral equation in
Eq.~(\ref{BHvK})~\cite{Bedaque:2002yg}. The need for a three-body contact interaction at leading order
implies that---at fixed $a$---all observables in the three-body sector
can be described by one-parameter correlations.  For
example, at fixed scattering length, particle-dimer scattering is
correlated with the three-body binding energy. Mapping out such a
correlation corresponds to varying $\kappa_*$, or $H_0$, keeping
$C_0$ fixed. One can, however, also vary the two-body contact
interaction and thus change $a$, keeping the three-body
counterterm fixed. This is what is expected to happen in atomic systems
close to a Feshbach resonance. There, a small variation in the
magnetic field changes the scattering length dramatically, but is not
expected to affect the short-distance physics that determines the
value of $H_0$.

Since $H_0$ (and hence $\kappa_*$) does not change as $a$ is varied we
have straightforward relations between the scattering lengths at which
certain features in atomic recombination are observed.  For example,
the scattering length $a_*$ at which the trimer binding energy crosses
the atom-dimer threshold is related to the binding momentum of the
corresponding trimer at unitarity by $a_*=0.071 /\kappa_*$~\cite{Braaten:2004rn}.


Such relations can be listed for all relevant quantities and
then lead to results such as Eq.~(\ref{eq:ratio}). These equations
will, however, be modified in presence of a finite effective
range. In the next section we consider these effects.
\section{Beyond Leading Order}
\begin{figure}
\centerline{\includegraphics[width=9cm,angle=0,clip=true]{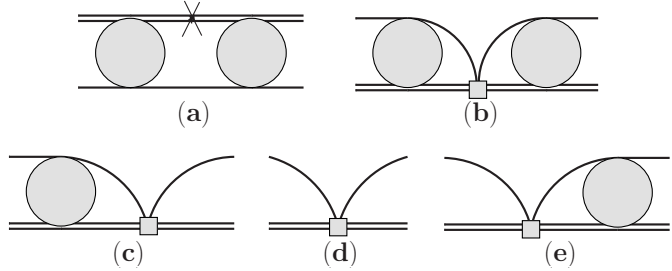}}
\caption{Diagrams for the NLO amplitude of atom-dimer scattering. The
 crossed double line denotes an insertion of the range propagator, ${\mathcal D}_1(|{\bf q}|;q_0)$, 
 the shaded blob denotes an insertion of the LO three-body amplitude, ${\cal A}_0$,
 and the square denotes an insertion of the NLO three-body force, $H_1$.
}
\label{fig:diagrams}
\end{figure}

To include higher-order corrections we follow and extend\footnote{In
  Ref.~\cite{Hammer:2001gh} one term stemming from two-body scattering
  was neglected in the analysis. It will be taken into account here. }
Hammer \& Mehen's analysis~\cite{Hammer:2001gh} and include the
operator associated with the two-body effective range, $r_s$, perturbatively in the
calculation of observables. This treatment is valid provided that $|r_s| \sim \ell \ll a$. To calculate effects $\sim r_s$  we consider the piece of the
two-particle propagator that contains one insertion of effects beyond
LO:
\begin{equation}
 \label{eq:D1}
 \mathcal{D}_1(|\qvec|;q_0)=\frac{r_s}{2}
\frac{1/a+\sqrt{-m q_0+\qvec^2/4-i\epsilon}}
{-1/a+\sqrt{-m q_0+\qvec^2/4-i\epsilon}}~.
\end{equation}
Diagram (a) of Fig.~\ref{fig:diagrams} then
corresponds to a single insertion of the propagator (\ref{eq:D1}) in
the three-body system. It was shown in Ref. \cite{Hammer:2001gh} that
the corresponding integral displays a divergence that has to be
canceled with an energy-independent three-body force
$H_1(\Lambda)$. Diagrams (b)--(e) in Fig. 1 have to be evaluated for this purpose. 

When this is done we obtain:
\begin{eqnarray}
 \label{eq:amplitude_nlo}
\nonumber
 \mathcal{A}_{1}(k,k)&=&\frac{r_s}{a} \mathcal{A}_0(k,k)
+\int_0^\Lambda\hbox{d}q\,q^2\frac{ a}{4 \pi^2 }
\mathcal{D}_1(q;E)
\mathcal{A}_0^2(k,q)
\nonumber
\\
&&\hspace{-20mm}+\frac{a H_1(\Lambda)}{\pi^3 \Lambda^2}
\left[\frac{4\pi^2}{a}+\int_0^\Lambda\hbox{d}q\,q^2
\frac{\mathcal{A}_0(k,q)}{-1/a+\sqrt{3 q^2/4-mE-i\epsilon}}\right]^2.
\end{eqnarray}
where the first term comes from the NLO renormalization of the dimer field. 
Inserting Eq.~(\ref{eq:1}) into the integral in the first line of this
expression makes it clear that the term $\sim H_1$ on the second line
has to absorb two independent divergence structures, one of which is
proportional to $1/a$.  Therefore we must write
\begin{equation}
\label{eq:3}
 H_1(\Lambda)= H_{10}(\Lambda) + \frac{1}{a} H_{11}(\Lambda)~.
\end{equation}
Considering only fixed scattering length does not introduce an additional
three-body parameter, since the total $H_1(\Lambda)$ is
fixed by the renormalization condition that the NLO amplitude has
zero effect on the observable used for renormalization
at leading order. However, if we desire predictions at NLO for
observables as a function of $a$ we need to know the relative
sizes of $H_{11}$ and $H_{10}$.

The existence of such an $a$-dependent three-body force can be
understood by considering the Lagrangian shown in
Eq.~(\ref{eq:5}). This Lagrangian is constructed by writing down all
terms allowed by the symmetries of the system. This includes terms
proportional to powers of the small scales inherent to the problem: in
our case, $1/a$. Beyond LO the three-body term in Eq.~(\ref{eq:5})
must be augmented to:
\begin{equation}
 D_0 (\psi^\dagger \psi)^3 +\frac{D_{11}}{a} (\psi^\dagger \psi)^3
+\frac{D_{22}}{a^2}(\psi^\dagger \psi)^3+\ldots.
\end{equation}

This exemplifies the benefit of any EFT: {\it all} operator structures
allowed by the underlying symmetries will contribute at an order in
the small-parameter expansion that can be determined by careful
analysis. It furthermore relates experiments with ultracold atoms to a
problem encountered in the application of $\chi$PT to nuclear
physics. In this description of low-energy nuclear processes using
only baryonic and pionic degrees of freedom, the small scale analogous
to $1/a$ is the pion mass $m_\pi$, and operators proportional to
powers of the pion mass are a mandatory part of expressions for
low-energy nuclear observables. For example, at $\mathcal{O}(m_\pi^2)$, the
nucleon-nucleon ($NN$) potential calculated in $\chi$PT can be written
as
\begin{equation}
V=[C_{S} + D_{S} m_\pi^2] + [C_{T} + D_{T} m_\pi^2] \sigma_1 \cdot \sigma_2 + V_{OPE} + V_{TPE}^{(2)}~,
\end{equation}
where $V_{OPE}$ is one-pion exchange and $V_{TPE}^{(2)}$ is the
``leading'' two-pion exchange contribution to the $NN$ interaction.
Both $C_{S/T}$ and $D_{S/T}$ are required for consistent
renormalization~\cite{Kaplan:1996xu}. The relative values of these two are not required
for the calculation of observables measured in the laboratory, i.e. at
fixed $m_\pi$. However, as lattice QCD attempts to predict and
explain few-hadron properties, the determination of pion-mass
dependent coefficients becomes a crucial feature of future progress
\cite{Beane:2008dv}.

\section{Three-Body Recombination}
\label{sec:three-body-recomb}
\subsection{The Lithium-7 System}
\begin{figure}
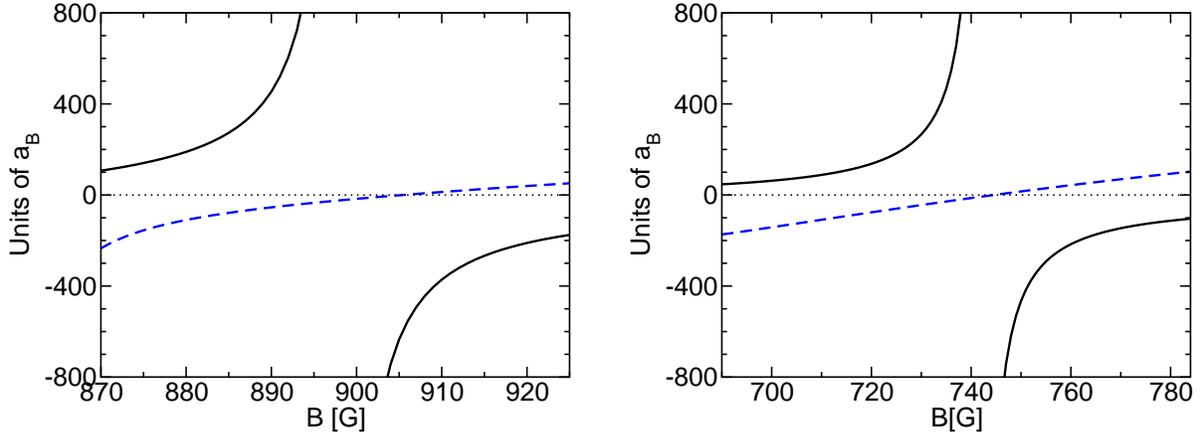

\centerline{\includegraphics[width=7.5cm,angle=0,clip=true]{a1-r1-B.eps}
\hspace{0.5cm} 
\includegraphics[width=7.5cm,angle=0,clip=true]{a2-r2-B.eps}}
\caption{(Color online) Magnetic-field dependence of the scattering length
and effective range in the hyperfine states relevant to the
experiments by Gross {\it et al.} (left panel) and Pollack {\it et al.}
(right panel). The solid lines denote the scattering length and the
dashed lines denote the effective range.
\label{fig:scattlengths}}
\end{figure}

The new three-body counterterm $H_{11}$ ($\equiv D_{11}$) is only
relevant if systems with a variable scattering length are
considered. Experiments with ultracold atoms are therefore the ideal
place to explore its impact.  Here we focus on $^7$Li, where we have
information on how both the scattering length and the effective range
vary with magnetic field, $B$. The dependence of $a$ and $r_s$ on the
magnetic field as presented in Ref.~\cite{Gross:2009} is shown in
Fig.~\ref{fig:scattlengths} for two different hyperfine
states~\cite{Kokkelmans}.

We will use our EFT, with this two-body input, to discuss two recent
experiments which measured three-body recombination rates in systems
of ${}^7$Li atoms.  First, Gross {\it et al.} have measured the
three-body recombination of atoms in the $| F=1\; m_F=0\rangle$
hyperfine state \cite{Gross:2009}. They found a recombination minimum
at $a>0$ and a recombination maximum at $a<0$ whose relative positions
are quite well described by universal ($r_s=0$) predictions. Second,
Pollack {\it et al.} have measured the recombination rate of atoms in
the ${\textstyle | F=1\; m_F=1\rangle}$ state and found several
recombination features associated with few-body universality
\cite{pollack:2009}. In these data there was a systematic deviation by
a factor of two from the universal prediction in the ratios of
features on the positive and negative scattering-length sides of the Feshbach resonance.

Using the diagrammatics discussed above we perform an EFT calculation of the
three-body recombination rate into dimers with binding energy $\sim 1/m a^2$,
for positive scattering length. Since our calculation does not include dimers
bound by $\sim 1/m \ell^2$, we cannot compute the recombination rate for
negative scattering length. We can, however, determine the position of
recombination maxima at $a<0$, by calculating the scattering lengths for which
the binding energy of a trimer becomes zero.

\begin{figure}
\centerline{\includegraphics[width=9cm,angle=0,clip=true]{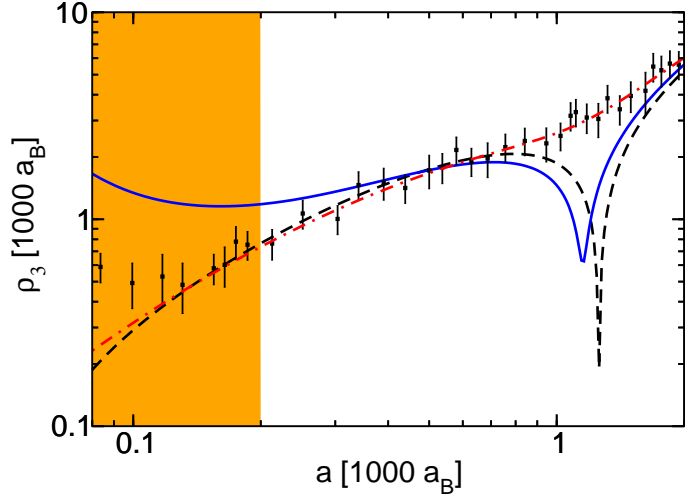}}
\caption{(Color online) The three-body recombination length of $^7$Li in the
  $| F=1\; m_F=0 \rangle$ hyperfine state as a function of the scattering
  length. The dashed line denotes the LO result where the three-body parameter
  is determined by the recombination maximum that has been measured
  experimentally in Ref.~\cite{Gross:2009}. The dot-dashed curve is
    the LO result including deep-dimer effects with fitting parameters as in
    Ref.~\cite{Gross:2009}. The solid line denotes our NLO result (which does not include deep dimers)}.
\label{fig:lithium-01}
\end{figure}

The recombination length, $\rho_3$, is obtained from the LO and NLO
elastic atom-dimer scattering amplitudes, ${\cal A}_0$ and ${\cal A}_1$, via
(in units where $\hbar=1$):
\begin{equation}
  \rho_3=2 \sqrt{\frac{1}{\gamma} \left|\left({\cal A}_0 + {\cal A}_1\right)\left(0,\frac{2 \gamma}{\sqrt{3}};0\right)\right|},
\end{equation}
with $\gamma=\sqrt{m B_2}$, and $B_2$ the binding energy of the atom-atom
dimer ($\gamma=1/a$ at LO). Figure \ref{fig:lithium-01} shows our results for
$\rho_3$ as a function of the scattering length $a$ for the $|
F=1\; m_F=0\rangle$ hyperfine state.  The dashed line denotes the LO result
renormalized to the recombination maximum $a_*'$ on the
$a<0$ side. This LO calculation predicts $a_{*0} > 1200 a_B$,
  c.f. the measured $a_{*0} \approx 1160 a_B$.  Because, as discussed above,
  consistent renormalization at NLO requires us to choose two different
  three-body observables to fix the counterterms $H_{00} + H_{10}$ and
  $H_{11}$, we can describe both the position of this recombination minimum
  {\it and} that of the observed maximum at $a < 0$.  The solid line displays
  that NLO result, renormalized to $a_*'$ and the recombination minimum
  $a_{*0}$ determined by Gross et al. \cite{Gross:2009}. The shaded area
  denotes the region where $|r_s|/a>0.5$ and convergence of the EFT expansion
  is expected to be slow. The squares give the experimental data with the
  corresponding errors. We emphasize that we did not strive for a detailed
  reproduction of the experimental data here, since our predictions account
  for neither the effects of deep dimers nor those of finite temperature. The
  inclusion of such physics improves the overall agreement between experiment
  and theory. To demonstrate this we also present the LO result with
  deep-dimer effects included: it is the dot-dashed line in
  Fig.~\ref{fig:lithium-01}.  This shows that the positions of the loss
  features which we are focusing on in this letter are not affected by
  recombination into deep dimers. They are also not affected by finite
  temperature~\cite{Braaten:2006qx}.

It is gratifying that we can explain Ref.~\cite{Gross:2009}'s
  measurements of both $a'$ and $a_{*0}$ and still obtain results consistent
  with the EFT expansion, but neither of these is an NLO prediction of the
  EFT. We now use our NLO calculation to predict the scattering length at
  which the atom-dimer resonance occurs, $a_*$, i.e. the atom-atom scattering
  length at which the atom-dimer scattering length diverges. The LO result
for $a_*$ is $a_*=-1.03 a_*'$, which puts $a_*$ in the region where $|r_s|/a
\sim 0.3$.  Therefore effective-range corrections to this observable can be
large. The result for $a_*$ thus depends strongly on the LO
amplitude that is employed in this calculation and so on the
observable that was used at LO as three-body input. We obtain
\begin{equation}
a_*=(271 - 105 + \ldots) a_B
\label{eq:astar1}
\end{equation}
when $a_*'$ is used at LO and $a_{*0}$ is used in addition at
NLO. Alternatively, we find
\begin{equation}
a_*=(257 - 4 + \ldots) a_B
\label{eq:astar2}
\end{equation}
when $a_{*0}$ is used at LO and $a_{*}'$ is used in addition at
NLO. In either case, our results suggest that $a_*$ is shifted to
smaller values than the LO prediction. The significant difference
between the NLO prediction in the two different renormalization
schemes provides an estimate of effects at NNLO and beyond. In fact,
we can expect that NNLO corrections are larger for the result
(\ref{eq:astar1}), since the LO renormalization point is further from
the unitary limit there. But, to encompass both (\ref{eq:astar1}) and
(\ref{eq:astar2}), as well as provide an uncertainty based on
conservative estimates of NNLO effects, we quote:
\begin{equation}
a_*=(210 \pm 44)  a_B.
\end{equation}

\begin{figure}
\centerline{\includegraphics[width=9cm,angle=0,clip=true]{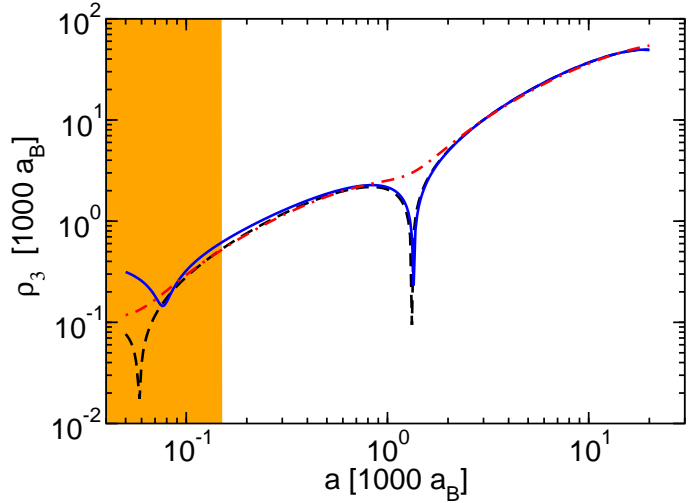}}
\caption{(Color online) The three-body recombination length of $^7$Li in the
  $| F=1\; m_F=1\rangle$ hyperfine state as a function of the scattering
  length. The dashed line is the LO result, with the three-body
    parameter determined by the recombination maximum on the $a<0$ side at
    $a_*'^{(1)}=-6301 a_B$ . The dot-dashed curve is the LO result
    including deep-dimer effects with the deep-dimer fitting parameter of
    Ref.~\cite{pollack:2009}. The solid line is our NLO result with the
  second three-body parameter adjusted to the second recombination maximum at
  $a_*'^{(2)}=-298 a_B$.}
\label{fig:lithium-11}
\end{figure}

We have performed a similar analysis for the results obtained by the
Rice group in Ref.~\cite{pollack:2009}. In Fig.~\ref{fig:lithium-11} we show our results for the
recombination length of ${}^7$Li atoms in the $| F=1\; m_F=1\rangle$ state as a function of the scattering length. The LO
result (dashed line) is renormalized to the first recombination
maximum $a_*'^{(1)}=-6301 a_B$ on the $a<0$ side.
The solid line shows the NLO result that is additionally renormalized to the
second measured recombination maximum $a_*'^{(2)}=-298 a_B$.
Having adjusted both three-body parameters, we can use them to
calculate other observables such as 
\begin{equation}
  \label{eq:6}
  \kappa_*=(0.242 \pm 0.030)\cdot 10^{-3} a_B^{-1},
\end{equation}
and the position of the atom-dimer resonance
\begin{equation}
  \label{eq:8}
  a_*=(355.8 \pm 55.5)  a_B~.
\end{equation}
In the latter case a shift upward from the LO prediction $a_*^{(LO)}=295 a_B$ is seen.
We can also predict the position of the second recombination
minimum in Fig.~\ref{fig:lithium-11} to be:
\begin{equation}
  \label{eq:9}
  a_{*0}=(1348 \pm 151 )a_B~.
\end{equation}
The errors in Eqs.~(\ref{eq:6})--(\ref{eq:9}) were obtained by
propagating the systematic and statistical errors quoted in
Ref.~\cite{pollack:2009}. The shift of the central value in
(\ref{eq:9}) from the LO result is only $26 a_B$, and so range effects
$\sim r_s^2$ and higher are certainly much smaller than the
uncertainty due to the experimental input.
The EFT result (\ref{eq:9}) therefore
disagrees with the experimental result obtained in Ref.~\cite{pollack:2009}
\begin{equation}
  \label{eq:10}
  a_{*0}=(2676 \pm 67 \pm 128 )a_B~,
\end{equation}
by a factor of two, even after range corrections are included. (In
Eq.~(\ref{eq:10}) the first error is statistical and the second due to
a systematic uncertainty in the determination of the atom-atom
scattering length.)  Effects due to the finite effective range 
are therefore not responsible for the disagreement between
the data of Ref.~\cite{pollack:2009} pertaining to different sides of
the Feshbach resonance and the predictions of universality.

\section{Conclusion}
\label{sec:conclusion}
We have shown that at next-to-leading order in the EFT for systems
with large scattering length an additional three-body input is
required to describe physical observables as a function of the
two-body scattering length. This parameter is needed for a complete 
NLO description of such observables if $|r_s| \sim \ell$. It encodes
the effect of dynamics at distances comparable to the range of the
underlying two-body force on the scattering-length dependence of
quantities in the three-body system. The necessity of
this operator for renormalization exemplifies the general tenet of all
EFTs, that all operators which are allowed by the
symmetries
of the system will appear in the EFT Lagrangian. An expansion of the
short-distance physics around the unitary limit (i.e. in powers of
$1/a$) makes clear that such an operator should be present at NLO in
the short-range EFT. In this respect $1/a$ in this EFT plays a similar
role to that of the pion mass in $\chi$PT.

Here we have analyzed the typical situation where the interaction range
and the two-body effective range are of the same size: $\ell \sim |r_s|$.
We expect the impact of our additional three-body counterterm
to be smaller if $\ell \ll |r_s|$, i.e. the effective
range is unnaturally large. This occurs, for example, close to a
narrow Feshbach resonance. In that case the results of
Ref. \cite{Platter:2008cx} should hold. It is possible that
the iteration of $r_s$ to all orders can be systematically
justified in such systems. In that case a three-body counterterm
would not be required at LO of the EFT calculation~\cite{petrov-reco}
and corrections to that ``resummed LO" would scale as $r_s/a$
and $k r_s$.

We have shown that our approach can be applied to
three-body recombination provided that the effective range is
known. In the case of $^7$Li we calculated the shift in recombination
features due to an effective range that is a function of the magnetic
field. We found that deviations from universal predictions in the
Bar-Ilan experiment can be explained by these effects but that they
are not sufficient to explain the inconsistencies across the resonance
in the experiment carried out by the Rice group. 

The data of Ref.~\cite{pollack:2009} is not consistent with $\sim r_s$ corrections to universality
 in spite of the presence of the extra three-body parameter at that order in our
 calculation. It has been suggested that this apparent violation of universality
is a result of the conditions in the Rice experiment. For
positive $a$ the Rice group used a BEC, while for negative $a$ they used a thermal
gas. In contrast, the Bar-Ilan group recently repeated the Rice $^7$Li experiment in the $| F=1,\; m_F=1\rangle$ state using a
thermal gas on both sides of the Feshbach resonance. They found that features across the resonance were related by universality~\cite{Gross:2010}.

Our results show that effects proportional to $r_s$ correct the
universal relations displayed, e.g. in Ref.~\cite{Braaten:2004rn} in a
way that improves the agreement with data in the case of ${}^7$Li
atoms~\cite{Gross:2009}. It would be very interesting to apply this
framework to data on three-body recombination of ${}^{133}$Cs
atoms. The existing data there, though, is rather different, since the
features observed in experiments with ${}^{133}$Cs at $a>0$ and $a<0$
are connected through the region where $a=0$~\cite{kraemer:2006}. In
the vicinity of this scattering-length zero $r_s/a$ diverges, and
higher-order corrections to the short-range EFT cannot be reliably
calculated. The connection between $a_*$~\cite{knoop:2009}, $a_*'$,
and recombination minima at $a > 1000 a_B$ can, however, be addressed
within EFT.
Given information on how the ${}^{133}$Cs two-body
scattering length and effective range vary with magnetic field,
we can make quantitative predictions for the impact of the range on
three-body recombination features seen in experiments with cold gases
of Cesium atoms.

\acknowledgments
  We thank E.~Braaten for helpful discussions. We acknowledge the INT
  program ``Simulations and Symmetries: Cold Atoms, QCD, and
  Few-hadron Systems", during which this work was completed. This
  research was supported in part by the DOE under grants
  DE-FG02-00ER41132, DE-FG02-93ER40756 and DE-FC02-07ER41457, by the
  NSF under grant PHY-0653312, and by the Mercator programme of the
  DFG.

\end{document}